\documentclass[a4paper,11pt]{article}
\usepackage{jinstpub} 
\usepackage{lineno}
\usepackage[acronym]{glossaries}
\usepackage[separate-uncertainty=true,per-mode=symbol]{siunitx}
\usepackage{tikz}

\usetikzlibrary{arrows.meta}

\glsdisablehyper
\newacronym{adu}{ADCu}{Analog-to-Digital Converter units}
\newacronym{asic}{ASIC}{Application-Specific Integrated Circuit}
\newacronym{apd}{APD}{Avalanche PhotoDiode}
\newacronym{enc}{ENC}{Equivalent Noise Charge}
\newacronym{ir}{IR}{InfraRed}
\newacronym{fwhm}{FWHM}{Full Width at Half Maximum}
\newacronym{snr}{SNR}{Signal-to-Noise Ratio}
\newacronym{qe}{QE}{Quantum Efficiency}
\newacronym{lgad}{LGAD}{Low Gain Avalanche Diode}
\newacronym{ilgad}{iLGAD}{inverse Low Gain Avalanche Diode}
\newacronym{pxs}{PXS}{Plasma X-ray Source}
\newacronym{rzp}{RZP}{Reflection Zone Plate}
\newacronym{rms}{RMS}{Root Mean Square}
\newacronym{msaxs}{mSAXS}{magnetic Small Angle X-ray Scattering}
\newacronym{psi}{PSI}{Paul Scherrer Institute}
\newacronym{tew}{TEW}{Thin Entrance Window}
\newacronym{fbk}{FBK}{Fondazione Bruno Kessler}
\newacronym{mbi}{MBI}{Max Born Institute}
\newacronym{xmcd}{XMCD}{X-ray Magnetic Circular Dichroism}

\newcommand{\moench}{MÖNCH}

\newcommand{\felthree}{\qty{707}{\eV}}

\newcommand{\gdmfive}{\qty{1190}{\eV}}
\newcommand{\dymfour}{\qty{1333}{\eV}}
\newcommand{\dymfive}{\qty{1292}{\eV}}

\DeclareSIUnit\adu{ADCu}
\DeclareSIUnit\electron{\text{e\ensuremath{^{-}}}}
\DeclareSIUnit\pixel{pixel}
\DeclareSIUnit\px{px}

\newcommand{\timingpanel}[6]{
\begin{tikzpicture}[
    >=Latex,
    trig/.style={-{Latex[length=2.5mm]}, thick, red!40!black},
    xray/.style={-{Latex[length=2.5mm]}, thick, blue},
    wave/.style={-{Latex[length=2.5mm]}, thick},
    decay/.style={thick, red}
]
    \node at (-1.4,1.75) {#1};
    \def\lowy{0}
    \def\highy{0.9}
    \draw[wave] (0,\lowy) -- (#3,\lowy) -- (#3,\highy) -- (#5,\highy) -- (#5,\lowy) --  (#6,\lowy) node[above] {\small $t$};
    \node[align=center] at (-0.8,0.45) {\small Exposure\\\small  window};
    \draw[trig] (#2,0) -- (#2,1.4);
    \node[align=center, above, text=red!40!black] at (#2,1.4) {\small Trigger in};
    \draw[xray] (#4,0) -- (#4,1.4);
    \node[align=center, above, text=blue] at (#4,1.4) {\small X-ray pulse};
    \draw[<->] (#2,0.45) -- (#3,0.45);
    \node at ({(#2+#3)/2},0.7) {\small $\tau$};
    \def\decaycurve##1{\highy*0.95*(1 - exp(-10*(##1-#4)))}
    \draw[decay] plot[domain=#4:#5, samples=60, smooth](\x, {\decaycurve{\x}});
    \draw[decay, densely dotted] plot[domain=#5:#6-0.2, samples=60, smooth](\x, {\decaycurve{\x}});    
    \node[red, right] at ({#6-0.7+0.03}, {\highy*0.95*(1 - exp(-5*(#6-0.5-#4))) + 0.35}) {\small $Q(t)$};
\end{tikzpicture}%
}

\title{Magnetic scattering experiments with an ultrashort pulsed soft X-ray source using iLGADs}

\author[a,*]{L. Lunin,\note[*]{Corresponding author.}}
\author[a]{S. Li,}
\author[a]{A. Bergamaschi,}
\author[b]{P. Wernicke,}
\author[b]{D. Schick,}
\author[b]{M. Mattern,}
\author[b]{B. Pfau,}
\author[c]{A. Bisht,}
\author[c]{M. Boscardin,}
\author[a]{M. Brückner,}
\author[a]{M. Carulla,}
\author[c]{M. Centis Vignali,}
\author[a]{R. Dinapoli,}
\author[a]{S. Ebner,}
\author[a]{S. Emiliani,}
\author[a]{K. Ferjaoui,}
\author[c]{F. Ficorella,}
\author[a]{E. Fröjdh,}
\author[a]{V. Gautam,}
\author[a]{D. Greiffenberg,}
\author[c]{O. Hammad Ali,}
\author[a]{S. Hasanaj,}
\author[a]{V. Hinger,}
\author[a]{M. Hürst,}
\author[a]{V. Kedych,}
\author[a]{T. King,}
\author[a]{C. Lopez-Cuenca,}
\author[a]{A. Mazzoleni,}
\author[a]{D. Mezza,}
\author[a]{K. Moustakas,}
\author[a]{A. Mozzanica,}
\author[a]{J. Mulvey,}
\author[a]{M. Müller,}
\author[c]{G. Paternoster,}
\author[c]{S. Ronchin,}
\author[a]{C. Ruder,}
\author[a]{B. Schmitt,}
\author[a]{S. Silletta,}
\author[a]{D. Thattil,}
\author[a]{X. Xie,}
\author[a]{J. Zhang}

\affiliation[a]{PSI Center for Photon Science, Paul Scherrer Institut,\\Forschungsstrasse 111, 5232 Villigen PSI, Switzerland}
\affiliation[b]{Max-Born-Institut für Nichtlineare Optik und Kurzzeitspektroskopie,\\Max-Born-Straße 2A, 12489 Berlin, Germany}
\affiliation[c]{Fondazione Bruno Kessler,\\Via Sommarive 18, 38123 Trento, Italy}

\emailAdd{leonid.lunin@psi.ch}

\abstract{\Gls{ilgad} sensors developed at the \gls{psi} in collaboration with the \gls{fbk} have demonstrated an excellent \gls{snr} and \gls{qe} in the soft X-ray range. In the current batch, single-photon detection is achievable at \qty{400}{\eV} with an \gls{snr} $>5$, while the \gls{qe} surpasses \qty{85}{\percent} at \qty{250}{\eV}. Combined with the high frame rates of CMOS readout ASIC, \gls{ilgad} sensors make the hybrid pixel detectors excellent candidates for time-resolved \gls{msaxs} experiments in the soft X-ray energy range.

In this contribution, we evaluated \Gls{ilgad} sensors bump-bonded to \qty{25}{\micro\meter} pitch charge-integrating \moench\ chips at a laser-driven soft X-ray source. During the experiments, we measured energy-dispersed radiation between \qty{600}{\eV} and \qty{1500}{\eV}. These measurements were performed for the \gls{ilgad} sensor, as well as for the standard silicon sensor. Using the high-precision electronic trigger of the X-ray source, we were able to synchronise the detector with the X-ray source delivering few-picosecond pulses and optimise the timing parameters for the acquisition.

Finally, we successfully performed the \gls{msaxs} experiment probing the magnetic domain pattern of a FeGd heterostructure, demonstrating better performance than a \moench\ detector with a standard silicon sensor. \Gls{ilgad} sensors achieved a significantly higher photon detection efficiency and, consequently, a higher sensitivity compared to standard silicon sensors under the same conditions, particularly for the lower-energy Fe L\textsubscript{3} edge.}
\keywords{Hybrid detectors, X-ray detectors, Instrumentation for synchrotron radiation accelerators}

\begin{document}
\maketitle
\flushbottom
\glsresetall
\section*{Introduction}
Soft X-ray experiments, especially those dealing with a weak observable photon signal, are limited by the capabilities of the state-of-art X-ray detectors. This applies both to the experiments yielding weak measurable signal including both at high flux radiation facilities as well as the lab based experiments with sources emitting orders of magnitude lower photon flux. In such experimental environments, high \gls{qe}, single-photon detection capabilities, high acquisition rates, and high duty cycles are essential for efficient data collection.

Recent studies in the development of \gls{ilgad} sensors for hybrid X-ray detectors have demonstrated a significant advantage in soft X-ray detection over standard silicon sensor technologies~\cite{andra2019,zhang2022}. Application of \gls{ilgad} sensors makes it possible to achieve a very high \gls{qe} for soft X-rays down to \qty{250}{\eV} while the intrinsic sensor signal amplification improves the \gls{snr}, enabling the detection of single photons down to \qty{550}{\eV} with a photon-counting \gls{asic}~\cite{baruffaldi2025} and \qty{400}{\eV} with a charge-integrating \gls{asic}~\cite{zhang2022}.

In this work we demonstrate the results of a pilot experimental application of the \gls{ilgad}-\moench\ frontend detector with a tabletop pulsed \gls{pxs}~\cite{borchert2023}. This includes the necessary preparatory steps for optimising the acquisition configuration. Further, the energy sensitivity capabilities in the energy range from \qtyrange{600}{1500}{\eV} are studied for both the \gls{ilgad} and standard silicon sensors. Finally, we provide a direct comparison of detection performance of the \gls{ilgad} and standard silicon sensors in a real-world \gls{msaxs} experiment.

\section{Magnetic Small Angle X-ray Scattering}
The study of nanometre-scale textures of magnetic materials, such as stripe and labyrinth domain patterns in ferromagnetic and ferrimagnetic thin films and multilayer systems, is of great interest for magnetisation dynamics research.

Soft X-ray-based techniques are particularly suitable for investigating magnetic properties of rare-earth and transition-metal elements, as the strong \gls{xmcd} effect at their absorption edges in soft X-ray can be exploited~\cite{fischer2015, vanderlaan1986} in combination with spatial resolution down to the nanometre scale.


For periodic magnetic structures, such as labyrinth domains, \gls{msaxs} reveals a ring-like diffraction pattern which directly reflects the average domain size as the ring radius. The integrated scattering intensity reflects the local magnetisation of domains, and thus enables time-resolved pump-probe experiments that follow ultrafast magnetisation dynamics on the femto- and picosecond timescales and nanometre length scale~\cite{bernard2026, lunin2025, pfau2012}.

\section{Soft X-ray source}
The measurements presented in this work were carried out using a laboratory-based \gls{pxs} as a source of soft X-ray radiation. The \gls{pxs} was developed and is stationed at the \gls{mbi} in Berlin~\cite{borchert2023}. This tabletop setup relies on a high-power pulsed \gls{ir} driving laser focused onto the surface of a tungsten target. Each sub-\qty{2}{\pico\second} \gls{ir} laser pulse generates a short-lived plasma that emits broadband sub-\qty{10}{\pico\second} radiation pulses over the full solid angle. A \gls{rzp} then captures and disperses a specific energy range of the soft X-ray radiation onto the X-ray detector sensor. An array of three \glspl{rzp} designed to cover the L-edges of Fe and M-edges of Gd and Dy is available for the experiments. High-precision hexapod mechanics allows reproducible positioning of the \glspl{rzp} between the measurements.

The number of photons in an individual pulse reaching the detector in the absence of a sample in the beam is of the order of \qtyrange{e3}{e4}{photon\per\eV} depending on the photon energy range. However, this number is significantly reduced by the sample absorption in spectroscopy or low cross-section in scattering experiments down to hundreds or dozens of photons per pulse in the relevant energy interval. This makes the use of a soft X-ray detector with high \gls{qe} and single-photon detection capability especially important for the present experimental setup.

The \gls{pxs} provides a high-precision electronic trigger for each X-ray pulse. With a repetition rate of \qty{100}{\hertz}, this allows a shot-to-shot acquisition scheme when a sufficiently high speed hybrid detector is used. The repetition rate of the setup could potentially be increased up to several \si{\kilo\hertz} and is currently limited by thin-disk laser and mechanical tungsten target.

\section{Detector frontend}
\subsection{\moench0.3 readout ASIC}
\label{text:moench-asic}
The hybrid detector architecture allows the readout chip, referred to as \gls{asic}, and the X-ray sensitive sensor to be produced in different technologies. The \moench0.3 charge-integrating \gls{asic} prototype with \numproduct{400x400} pixels and \qty{25}{\micro\meter} pixel pitch was used for the experiments~\cite{dinapoli2014}. The maximum frame rate is \qty{3}{\kilo\hertz} in a single network interface regime and  \qty{6}{\kilo\hertz} with two network interfaces. The \gls{asic} also allows parallel readout for dead-time-free acquisition.

The noise of the detector is quantified as \gls{enc}, i.e.\ is the input charge that would produce an output signal equal in magnitude to the RMS noise of the readout circuit~\cite{radeka1988}. For standard silicon sensors this ranges from about \qty{30}{\electron} for \qty{500}{\nano\second} exposure time to \qty{55}{\electron} for \qty{360}{\micro\second} with full duty cycle. Using sensors with internal amplification reduces the \gls{enc} almost proportionally to the gain factor~\cite{zhang2022}.

Due to the small pixel pitch, a significant fraction of the charge is shared among the neighbouring pixels. In order to reconstruct the photon induced charge and discriminate the detected photon signal from the background, \numproduct{3x3} pixel clusters were analysed with a photon-finding algorithm~\cite{cartier2014}.

\subsection{Inverse Low Gain Avalanche Diode sensors}
The \gls{lgad} sensor technology exploits an \gls{apd}-like mechanism~\cite{pellegrini2014}. Impact ionisation provides an amplification of the electric signal directly in the sensor structure through charge multiplication. Compared to \glspl{apd}, \glspl{lgad} provide moderate gain in range of \numrange{5}{20}, effectively limiting system shot noise. For \gls{ilgad}, the gain layer is uniformly implanted at the sensor backside (where the X-rays impinge) covering the entire pixel area, which provides a \qty{100}{\percent} fill factor.  The sensors used in this study have been developed in a collaboration between \gls{psi} and \gls{fbk}, focusing on high \gls{qe} and \gls{snr} for soft X-ray detection.

Standard sensors with an aluminised entrance window and non-optimised contact achieve a \gls{qe} lower than \qty{5}{\percent} for photon energy below \qty{400}{\eV}. To overcome this limitation, the \gls{tew} design was introduced, completely removing the metallisation layer and improving the charge collection efficiency by optimising the doping profile on the backside and introducing a custom thin surface passivation layer~\cite{zhang2022}.

A further consideration in sensor design optimisation concerns the limited penetration depth of soft X-ray photons in silicon, which extends only a few hundred nanometres beneath the surface. Because the impact ionisation factors for holes and electrons traversing the gain layer differ, the multiplication gain is reduced for X-rays absorbed near the surface, where hole-triggered multiplication prevails, relative to those absorbed within the bulk, where electron-triggered multiplication dominates, and the gain depends on the depth of absorption of the photons. To enhance the performance at low photon energies, a shallow and thin gain region was therefore adopted, maximising the proportion of incident photons absorbed beyond the gain layer.


The performance of \glspl{ilgad} is strongly temperature-dependent: the gain increases at low temperatures, while the leakage current decreases. Both properties allow to reduce the effective noise and improve the \gls{snr} at a low operating temperature~\cite{zhang2022}. Throughout all experiments, the \gls{ilgad} sensor was cooled and maintained at \qty{-20}{\celsius} for better single-photon detection capabilities. A bias voltage of \qty{300}{\volt} was applied to overdeplete the entire sensor volume.


The shallow gain layer \gls{ilgad} sensor with \gls{tew} bump-bonded to the \moench0.3 \gls{asic} was compared to the standard silicon sensor in the experiments presented in this manuscript. A comprehensive characterisation of this sensor can be found in~\cite{liguori2023}, where the sensor used in this work is referred to as W9.

\section{Trigger synchronisation optimisation}
In the context of a shot-to-shot acquisition scheme with a pulsed source, precise synchronisation of the detector exposure time window and X-ray pulse is essential for efficient photon signal collection.

\begin{figure}[ht!]
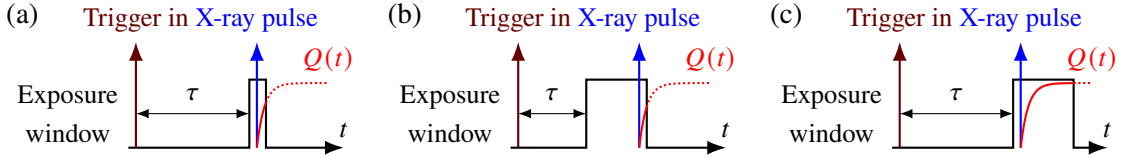

\centering
\timingpanel{(a)}{0.1}{1.6}{1.7}{1.82}{2.85}
\timingpanel{(b)}{0.1}{1.0}{1.7}{1.8}{2.85}
\timingpanel{(c)}{0.1}{1.6}{1.7}{2.4}{2.85}
    \caption{Schematic timing diagrams of acquisition showing: (a) too short exposure time - the charge $Q(t)$ is only partially transferred upon the exposure end, even though the X-ray pulse arrives right after the exposure start. (b) Despite the longer exposure time, the too short delay $\tau$ also results in partial charge integration. (c) The long exposure time and the optimal choice of delay $\tau$ maximises the integrated charge.}
    \label{fig:timing_scheme}
\end{figure}

The basic architecture of \moench0.3 pixel resembles that of \moench0.2 described in~\cite{dinapoli2014}. Exposure, charge integration and readout of the pixel are defined by the timed sequence of opening and closing of preamplifier reset and storage switches.
The exposure time window must be sufficiently long, and its start must be properly synchronised with the X-ray pulse arrival time. This is done with the adjustable delay $\tau$ between the X-ray source trigger signal and the start of exposure window, with the optimum being when the X-ray pulse arrives as early as possible after the exposure starts, as shown in \autoref{fig:timing_scheme}.

\begin{figure}[ht!]
    \centering
    \includegraphics[width=0.6\textwidth]{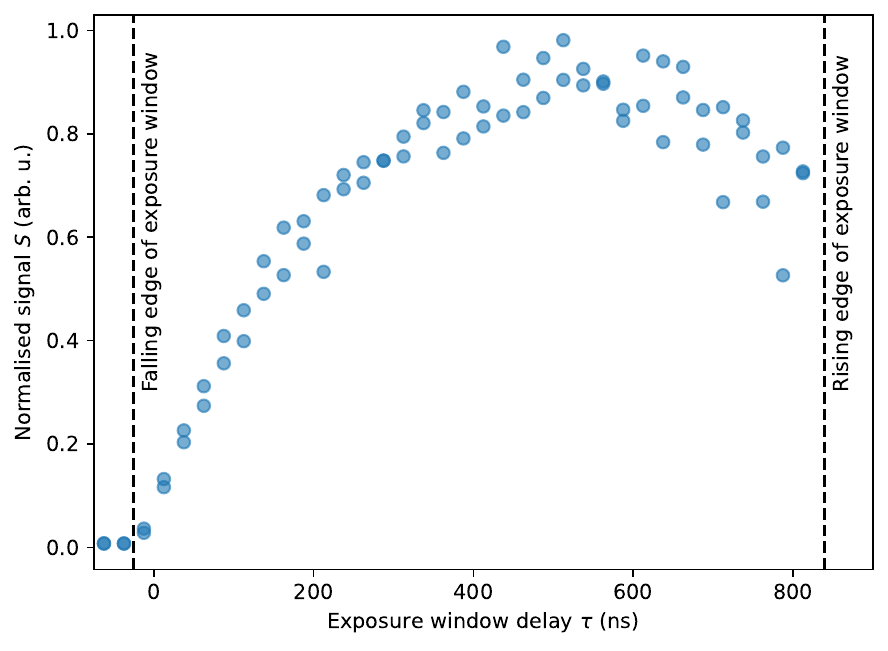}
    \caption{Signal $S$ integrated over the whole sensor as a function of the delay $\tau$, measured in two independent scans. A strong dependence of the detected intensity $S$ on the delay $\tau$ is observed, increasing with longer delays and reaching a plateau at approximately \qty{600}{\nano\second}.
    }
    \label{fig:delay_scans}
\end{figure}

\noindent
With the measurements of the signal $S$ integrated over the whole sensor area of the \moench\ detector as a function of the electronically adjustable delay $\tau$ between the external trigger and the exposure start, the optimal delay $\tau$ can be determined. Two scans of $S(\tau)$ are shown in \autoref{fig:delay_scans}.
As the noise contribution of the leakage current worsens with increasing exposure time, while the signal loss due to an incomplete signal settling reduces with increasing exposure time, $t =\qty{500}{\nano\second}$ and the corresponding delay maximising the signal were used for the measurements presented in this manuscript as an optimum between these two parameters.

\section{Detector characterisation}
\begin{figure}[ht!]
    \centering
    \includegraphics{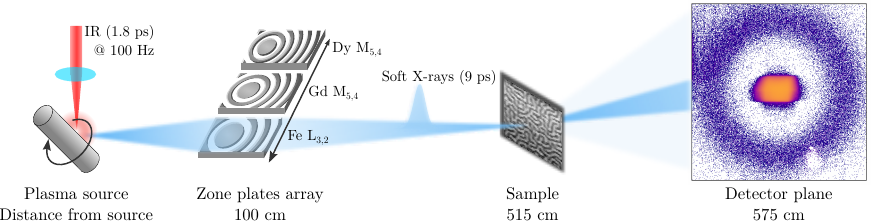}
    \caption{Schematic representation of experimental setup for the \gls{msaxs} experiments. For energy calibration of the setup and for energy sensitivity measurements, the magnetic sample is removed and a filter is placed into the beam.}
    \label{fig:setup_sketch}
\end{figure}

In \autoref{fig:setup_sketch} the principal scheme of the experimental setup is shown. A detailed technical description is given in~\cite{borchert2023}.%
\subsection{Spatial energy calibration}
\label{text:spatial_energy_calibration}
For the experiments, three different \glspl{rzp} were used with designed focus energies between Fe L\textsubscript{3,2}, as well as Gd and Dy M\textsubscript{5,4} edges. According to previous studies of \gls{rzp}~\cite{brzhezinskaya2013} and simulations performed with the \texttt{RAYX}~\cite{erdem2025} ray tracing simulation package, the photon energy $E$ is expected to be dispersed with a linear dependence along the vertical position $y$ on the sensor area.

\begin{figure}[ht!]
    \centering
    \includegraphics[width=0.9\textwidth]{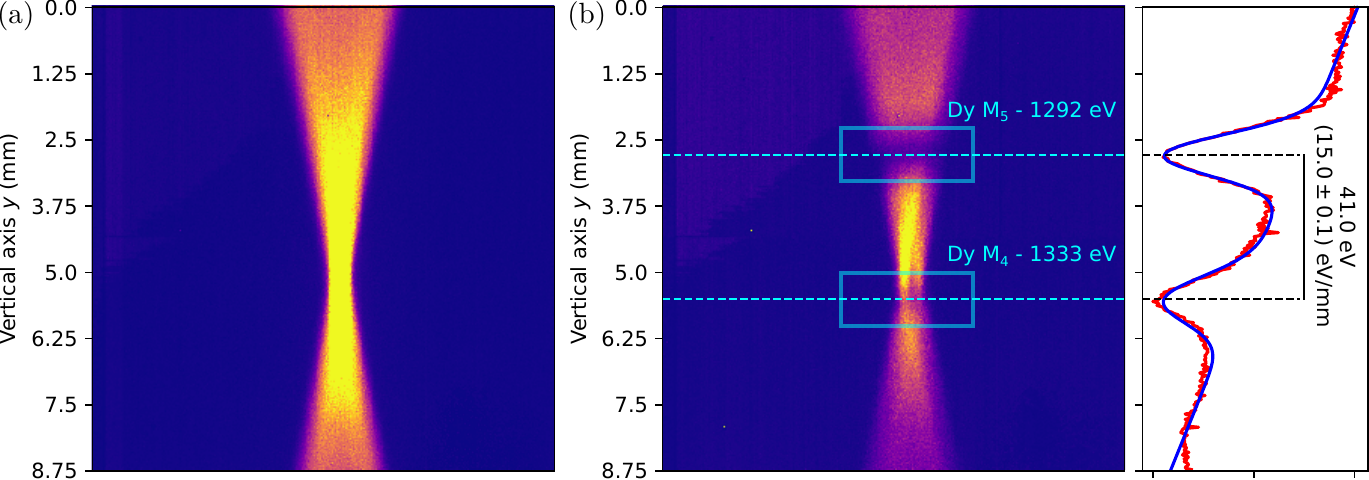}
    \caption{Reflection pattern of \gls{rzp} designed for Dy (a) without and (b) with a Dy filter placed in the beam. The projection of the spectra onto the vertical axis shows two horizontal absorption lines, corresponding to the Dy M\textsubscript{5} (\dymfive) and Dy M\textsubscript{4} (\dymfour) absorption edges. A fit function (blue) comprising the sum of two Gaussian peaks and a linear component is used to extract the positions of the absorption edges. Shown are the sums of \num{5000} individual frames.}
    \label{fig:dy_calibration}
\end{figure}

In order to perform energy calibrations for each of the \glspl{rzp}, free standing filters made of Fe, Gd, or Dy were placed into the beam, as shown in \autoref{fig:setup_sketch}. From the known photon energies of the absorption edges and the distance between the absorption lines on the detector, the dispersion (in \unit{\eV\per\milli\meter}) was calculated, as exemplarily shown for Dy in \autoref{fig:dy_calibration}.
This yielded \qty{15.0(1)}{\eV\per\milli\meter} for Dy and Gd \glspl{rzp} and \qty{7.9(1)}{\eV\per\milli\meter} for the Fe \gls{rzp}. This value depends only on the sensor position relative to \gls{rzp} and is the same for both \gls{ilgad} and standard silicon sensors, since both were mounted at the same position.

\subsection{Gain calibration}
\label{text:gain_calibration}
The \gls{msaxs} process is highly energy selective since the resonant condition for the scattered photons needs to be fulfilled. The resonant edge width of $\approx \qty{1.5}{\eV}$ \gls{fwhm} \cite{henke1993} can be considered negligible with respect to the photon energy. Another advantage of \gls{msaxs} data for energy calibration is that the scattered photons are spatially separated from the direct beam. Since the probability of scattering is very low, most of the detected photon events are the single-photon events required for the analysis.

\begin{figure}[ht!]
    \centering
    \includegraphics[width=0.6\textwidth]{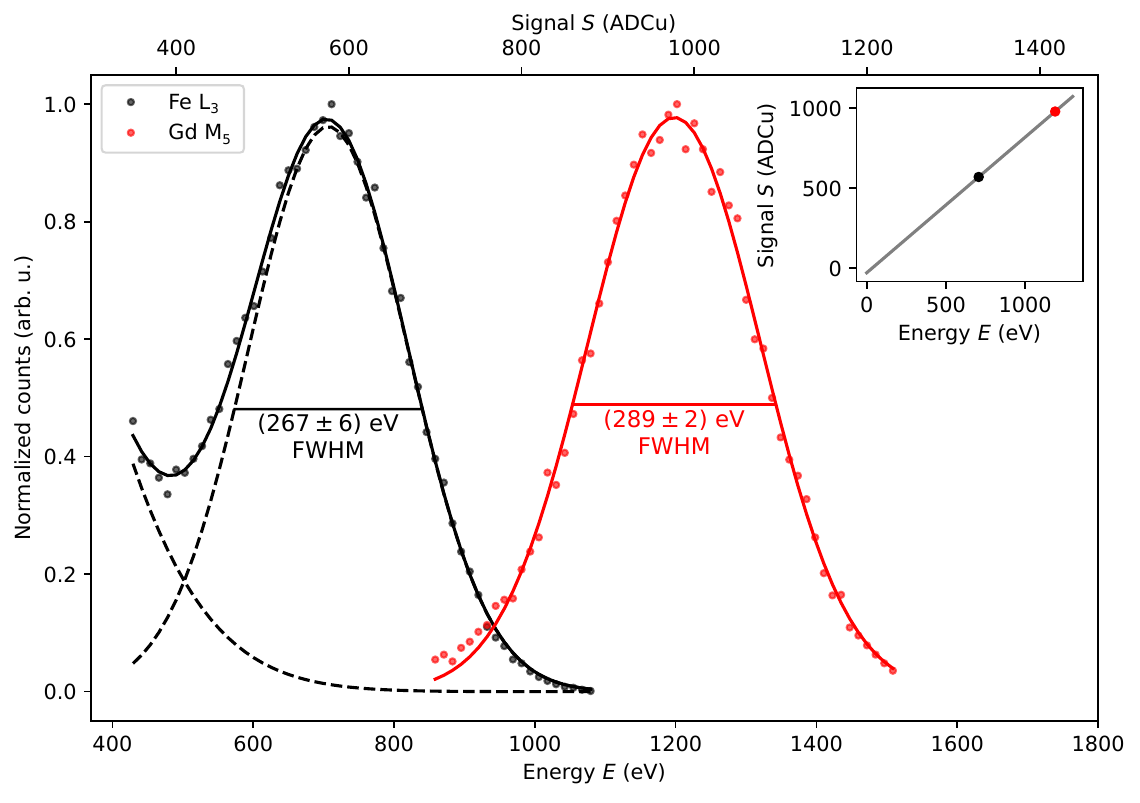}
    \caption{Distribution of (black) Fe L\textsubscript{3} and (red) Gd M\textsubscript{5} scattered cluster values accumulated over the whole \gls{ilgad} sensor. A fit function comprising the sum of two Gaussian peaks is used for the Fe L\textsubscript{3} and Gd M\textsubscript{5} spectrum. In the inset, the centres of the single-photon peaks \qty{570(1)}{\adu} and \qty{980(1)}{\adu} are extracted and plotted over known Fe L\textsubscript{3} and Gd M\textsubscript{5} photon energies, respectively. A linear fit function $kx+b$ is used to derive the slope $k=\qty{0.84}{\adu\per\eV}$ with a $b=\qty{-28}{\adu}$ offset.}
    \label{fig:w9_gain_calibration}
\end{figure}

Single-photon events of \gls{msaxs} were used to obtain the conversion factor from \gls{adu} to \unit{\eV} for the \gls{ilgad} sensor by calculating the ratio $g$ between the peak position of the spectrum $P$ in \gls{adu} and the X-ray photon energy.
Due to the low occupancy of each pixel, a mean conversion factor $g$ for the whole sensor area was derived. \autoref{fig:w9_gain_calibration} shows the spectra obtained for Gd M\textsubscript{5} and Fe L\textsubscript{3} photon energies for the \gls{ilgad} sensor.

The energy resolution for the standard silicon sensor cannot be estimated in the same way since the single-photon peaks at both the Fe L\textsubscript{3} and Gd M\textsubscript{5} photon energies cannot be clearly separated from the noise peak. Therefore, for the standard silicon sensor, \gls{adu} to \unit{\eV} mean conversion factor $g$ was derived previously in 2020 at the PTB metrology beamline at BESSY II.

With the known \gls{adu} to \unit{\eV} conversion factor $g$ for both of the sensor types, the \gls{ilgad} gain factor relative to the standard silicon \moench\ sensor is a factor \num{5.6}, which is due to the internal gain of the \gls{ilgad}. The observed \gls{fwhm} of the peaks indicates a weak dependence on the photon energy and is \qty{267(6)}{\eV} at the Fe L\textsubscript{3} and \qty{289(2)}{\eV} at the Gd M\textsubscript{5} edges due to the contribution of photons absorbed shallow before or within the gain layer, prevalent at lower energies and yielding to a lower gain.
\section{Energy sensitivity experiments}
Repetitive scans over the \gls{rzp} tilting angle i.e.\ the vertically dispersive axis of the beam were performed to verify the expected photon energy dependence on the vertical position. Measurements were performed for Fe, Gd and Dy \glspl{rzp}.

\begin{figure}[ht!]
    \centering
    \includegraphics[width=\linewidth]{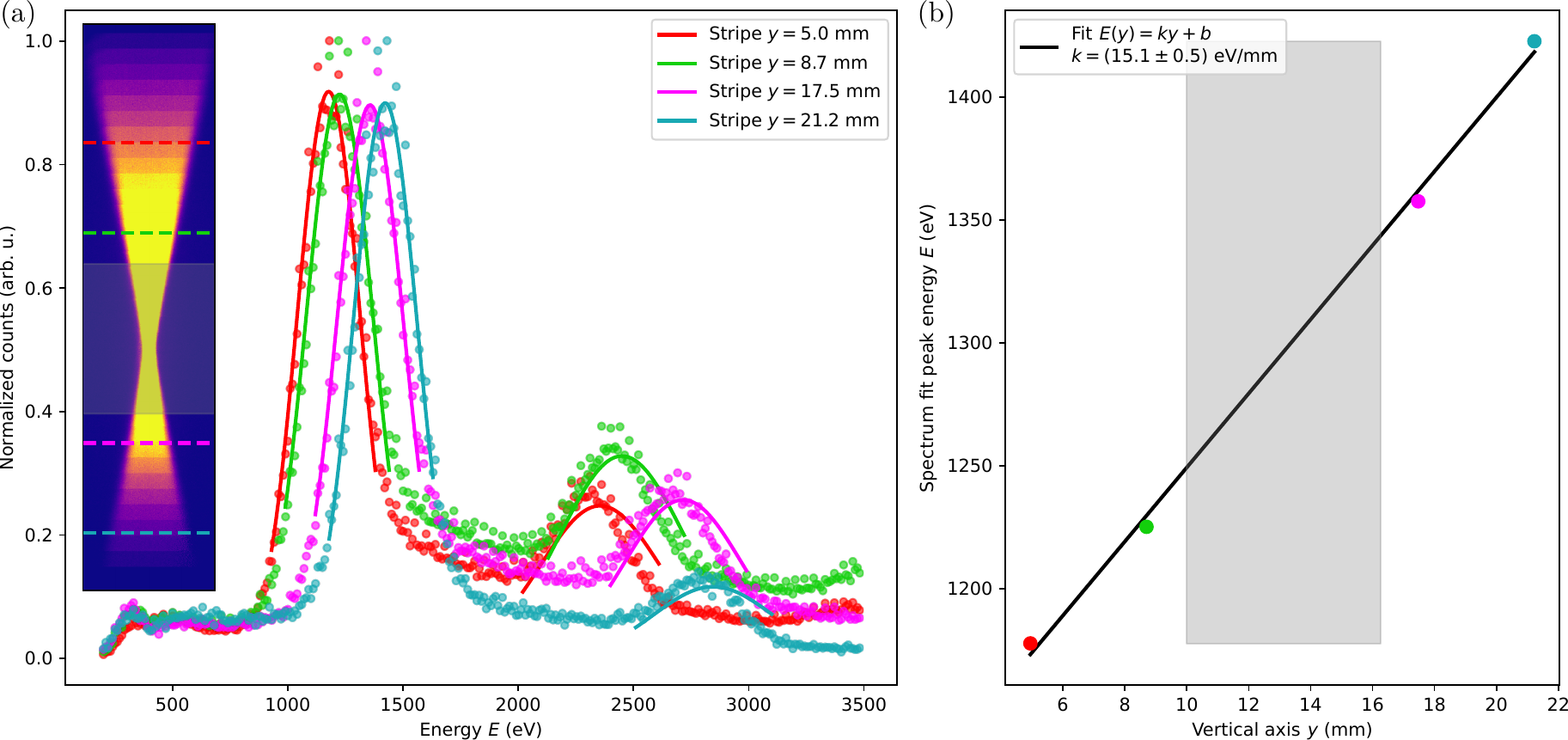}
    \caption{Spectra (a) of four horizontal stripes of width $W=\qty{0.15}{\milli\meter}$ (as shown in the inset in the same colours) of the \gls{ilgad} sensor are shown. A fit function comprising the sum of two Gaussian distributions, with the constraint that the mean value of the second Gaussian is twice that of the first is used. Peak positions corresponding to the single-photon energy are plotted against vertical position in (b), and a linear fit is applied to determine the energy slope along the vertical axis. The grey rectangle corresponds to the regions excluded from the analysis.}
    \label{fig:w9_dy_spectra_fit}
\end{figure}

The photon-finding algorithm was applied for each of the acquired frames to reconstruct the collected charge from the \numproduct{3x3} pixel cluster, as described in \autoref{text:moench-asic}. For each of the horizontal stripes of width $W$, a histogram of the photon energies was constructed and the single-photon energy peak was identified. Then the corresponding photon energy peaks positions for each of the horizontal stripes are plotted against the vertical coordinate $E(y)$. The \gls{rzp} focus region has much higher photon occupancy and is dominated by multi-photon events, making the region not eligible for analysis. This region is excluded from the collected data. The resulting $E(y)$ dependence was approximated with a linear fit and the energy slope in \unit{\eV\per\milli\meter} was determined. In \autoref{fig:w9_dy_spectra_fit} the steps described above are shown as an example for the Dy \gls{rzp}.

The energy slope of \qty{15.1(5)}{\eV\per\milli\meter} extracted with the linear fit is in agreement with the reference value previously determined in \autoref{text:spatial_energy_calibration}. The energy slope also matches the reference value for the same measurements for Gd and Fe \glspl{rzp} as well. The average peak width is \qty{332(10)}{\eV} \gls{fwhm}, which is close to the value for the photon energy, based on the previously determined peak \gls{fwhm} given in \autoref{text:gain_calibration}. The same analysis was not possible for the standard silicon sensor since no single-photon peak in a histogram can be clearly identified and separated from the noise, due to the too low \gls{snr}.



\section{Magnetic Small Angle X-ray Scattering experiments}
The sample system is an [Fe(\qty{0.4}{\nano\meter})|Gd(\qty{0.5}{\nano\meter})]$_{116}$ superlattice with antiferromagnetically coupled alternating Fe and Gd layers. The superlattice system exhibits magnetic stripe domains forming a labyrinth pattern with average period of \qty{300}{\nano\meter} in out-of-plane magnetisation.

Acquisition was performed for both a standard silicon sensor with \qty{500}{\nano\meter} Al coating and an \gls{ilgad} with a \qty{200}{\nano\meter} Al coating for light shielding. The same Peltier-cooled mounts were used for both the sensors, and the measurements were done at \qty{-20}{\celsius}. While the \gls{ilgad} sensor was supplied with a bias voltage of \qty{300}{\volt}, the standard silicon sensor was operated at \qty{180}{\volt}. For both sensors, \num{15000} frames were acquired at the two resonance energies with the highest scattering efficiency, namely Gd M\textsubscript{5} and Fe L\textsubscript{3}.

Then, the photon-finding algorithm was applied for each of the acquired frames to reconstruct the collected charge from the \numproduct{3x3} pixel cluster. A threshold of 3 times the cluster noise \gls{rms} $\sigma$ was applied to the sums of the cluster values to filter out most of the clusters consisting of background noise.

For the \gls{ilgad} sensor, the threshold was increased and set slightly below the single-photon peak edge of the corresponding photon energy, so the noise is reduced without loss of the truly detected photons. This corresponds to thresholds of about $\num{6.9}\sigma$ and $\num{7.8}\sigma$ for Fe L\textsubscript{3} and Gd M\textsubscript{5} photon energies, respectively, where $\sigma$ is the \gls{rms} noise of the \numproduct{3x3} cluster i.e.\ 3 of \gls{enc}.

For the standard silicon sensor, a $3\sigma$ threshold at Fe L\textsubscript{3} significantly reduces the number of detected photons, as the single-photon signal of Fe L\textsubscript{3} has a peak at $2.1\sigma$. Nevertheless, lowering the threshold significantly degrades the \gls{snr}, since the number of noise increases by orders of magnitude relative to the detected photon signal.

\begin{figure}[ht!]
    \centering
    \includegraphics[width=0.9\linewidth]{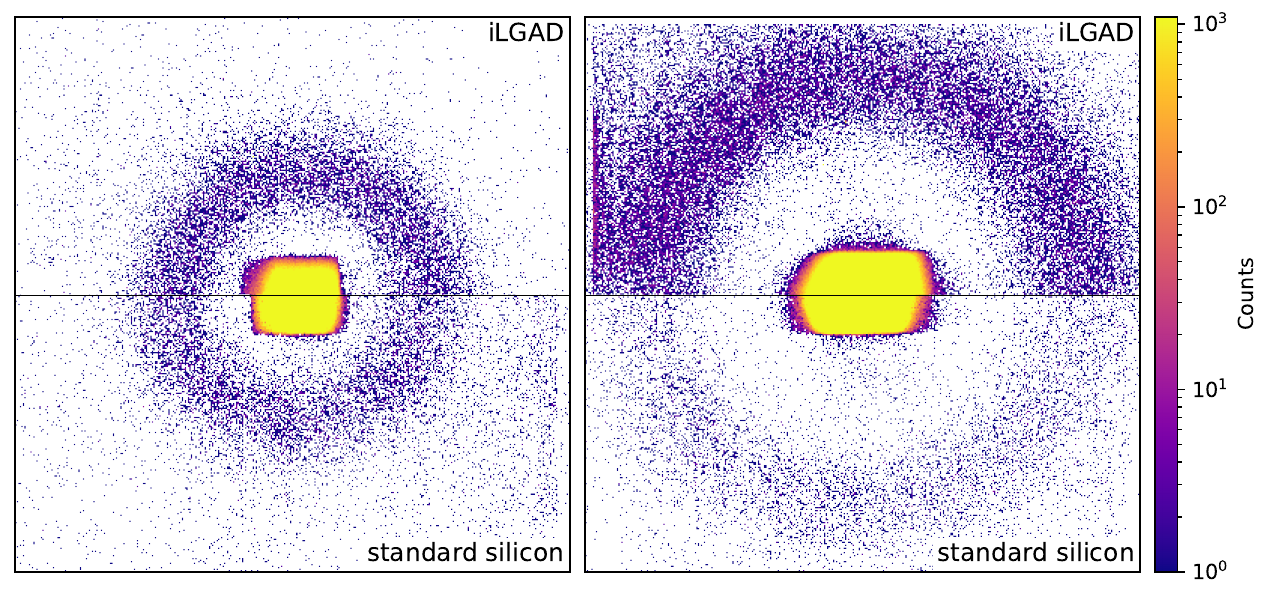}
    \caption{Composite image of \gls{msaxs} patterns acquired with (top half) \gls{ilgad} and (bottom half) standard silicon sensors. Scattering experiments were performed with the same sample with (left) Gd M\textsubscript{5} \gdmfive\ and (right) Fe L\textsubscript{3} \felthree\ resonant photon energies.}
    \label{fig:saxs_overview}
\end{figure}

\autoref{fig:saxs_overview} shows the direct comparison of the \gls{msaxs} patterns acquired with the \gls{ilgad} and standard silicon sensor for Gd M\textsubscript{5} and Fe L\textsubscript{3} photon energies.
The ring-like scattering pattern on the Fe L\textsubscript{3} edge exhibits a larger scattering angle $\alpha$ than that of the Gd M\textsubscript{5} edge, as the scattering angle $\alpha \propto q \lambda$ is proportional to the wavelength and inversely proportional to the photon energy $E$.

The \gls{ilgad} sensor demonstrated \num{5.1} and \num{1.3} improved detection efficiency over standard silicon sensor for scattering experiments on Fe L\textsubscript{3} and Gd M\textsubscript{5} resonant edges, respectively. The \gls{qe} of the standard silicon sensor are \qty{65}{\percent} and \qty{90}{\percent}, respectively, compared to \qty{80}{\percent} and almost \qty{100}{\percent} for the \gls{ilgad} sensor. While the \gls{qe} improvement factor of \num{1.11} for Gd M\textsubscript{5} photon energy nearly matches the detection efficiency gain, this effect alone is not sufficient to explain the difference in detection efficiency for Fe L\textsubscript{3} photon energy. The ratio of \gls{qe} for the \gls{ilgad} and standard silicon sensors for Fe L\textsubscript{3} yields a factor of $\approx \num{1.23}$ fewer photons to be detected with the standard silicon sensor. Moreover, since the threshold cut is higher than the single photon peak, only $\approx \qty{24}{\percent}$ of the photons will be detected.


The combined effect of \gls{qe} and threshold losses yields a factor of \num{5.2} fewer photons detected with a standard silicon sensor, which is in good agreement with the factor derived from the experiment.

\section{Summary and outlook}
The experiments presented in this manuscript demonstrate a significant improvement in single-photon detection efficiency in the soft X-ray range under real experimental conditions, combined with high energy sensitivity and an accurate energy response. At the Fe L\textsubscript{3} edge, an increase of detection efficiency by a factor of \num{5.1} was demonstrated in an actual \gls{msaxs} experiment. This opens the way for further soft X-ray scattering and imaging techniques for studying Fe alloys. Experiments at the oxygen K edge at \qty{543}{\eV}~\cite{mandziak2025,schneider2026} are also of particular interest for magnetism studies.

Higher gains are expected from the next production batch of \gls{ilgad} sensors with an optimised design. The sensors are expected to reach a gain of up to 20, compared to \num{5.6} for the sensor studied in this manuscript, and a \gls{qe} above \qty{80}{\percent} down to \qty{250}{\eV}~\cite{carulla2024}. Together with the parallel development of a full-scale \moench1.0\ \gls{asic} module with a larger sensor area (approx.\ \qtyproduct[product-units=single]{2x3}{\centi\meter}) and the same \qty{25}{\micro\meter} pixel pitch, this will offer a unique combination of characteristics for soft X-ray detection applications for imaging experiments.
\appendix


\bibliographystyle{JHEP}
\bibliography{biblio.bib}

\end{document}